\documentclass{aastex}
\usepackage{amssymb}
\usepackage{amsmath}
\usepackage{amsfonts}
\usepackage{graphicx}%
\begin{document}

\begin{center}
{\LARGE Electromagnetic thermal instability with momentum and energy exchange
between electrons and ions in galaxy clusters }

\bigskip

{\large Anatoly K. Nekrasov}

\bigskip

Institute of Physics of the Earth, Russian Academy of Sciences, 123995 Moscow, Russia

anatoli.nekrassov@t-online.de, anekrasov@ifz.ru

\bigskip

{\large ABSTRACT}
\end{center}

Thermal instability in an electron-ion magnetized plasma which is relevant in
the intragalactic medium (IGM) of galaxy clusters, solar corona, and other
two-component plasma objects is investigated. We apply the multicomponent
plasma approach when the dynamics of all species is considered separately
through the electric field perturbations. General expressions for the
dynamical variables obtained in this paper can be applied for a wide range of
astrophysical and laboratory plasmas also containing neutrals and dust grains.
We assume that background temperatures of electrons and ions are different and
include the energy exchange in the thermal equations for the electrons and
ions along with the collisional momentum exchange in the equations of motion.
We take into account the dependence of collision frequency on the density and
temperature perturbations. The cooling-heating functions are taken for both
electrons and ions. A condensation mode of thermal instability has been
studied in the fast sound speed limit. A new dispersion relation including the
different electron and ion cooling-heating functions and other effects
mentioned above has been derived and its simple solutions for growth rates in
the limiting cases have been found. We have shown that the perturbations have
an electromagnetic nature. The crucial role of the electric field perturbation
along the background magnetic field in the fast sound speed limit has been
demonstrated. We have found that at conditions under consideration, the
condensation must occur along the magnetic field while the transverse scale
sizes can be both larger and smaller than the longitudinal ones. The results
obtained can be useful for interpretation of observations of dense cold
regions in astrophysical objects.


\bigskip

\section{INTRODUCTION}

If a medium in a thermal equilibrium can become cooler due to radiation and
fluid contraction, it can get unstable, leading to formation of density
condensations with lower temperature than in the surrounding medium (Parker
1953; Field 1965). This instability called the thermal or
radiation-condensation one has been studied for more than five decades for
astrophysical objects and plasma physics applications (for reviews see, e.g.,
Meerson 1996; V\'{a}zquez-Semadeni et al. 2003; Elmegreen \& Scalo 2004; Cox
2005; Heiles \& Crutcher 2005). Many papers in astrophysical literature
considered the thermal instability in the neutral (Field 1965; Begelman \&
McKee 1990; Hennebelle \& P\'{e}rault 1999; Koyama \& Inutsuka 2000, 2002;
Burkert \& Lin 2000; Kritsuk \& Norman 2002; S\'{a}nchez-Salcedo et al. 2002;
Audit \& Hennebelle 2005; V\'{a}zquez-Semadeni et al. 2006; Hennebelle \&
Audit 2007 and references given above) and magnetized interstellar medium
(ISM; Field 1965; Hennebelle \& P\'{e}rault 2000; Stiele et al. 2006; Fukue \&
Kamaya 2007; Inoue \& Inutsuka 2008; Shadmehri et al. 2010), solar corona,
where prominences are formed (e.g., Field 1965; Nakagawa, 1970; Heyvaerts
1974; Mason \& Bessey 1983; Karpen et al. 1988), planetary nebulae (e.g.,
Field 1965), galaxy clusters and intragalactic medium (IGM; Field 1965;
Mathews \& Bregman 1978; Balbus \& Soker 1989; Loewenstein 1990; Balbus 1991;
Bogdanovi\'{c} et al. 2009; Parrish et al. 2009; Sharma et al. 2010). This
instability was studied for rotating (e.g., Field 1965; Nipoti 2010),
expanding (e.g., Field 1965; Gomez-Pelaez \& Moreno-Insertis 2002), and
dynamical systems with steady or time-dependent flows and nonstationary
background parameters (e.g., Mathews \& Bregman 1978; Balbus 1986; Balbus \&
Soker 1989; Burkert \& Lin 2000). The nonlinear stage of the thermal
instability resulting in formation of nonlinear structures (localized clouds)
in the ISM and solar prominences was investigated in (e.g., Trevisan \&
Ib\'{a}\~{n}ez 2000; S\'{a}nchez-Salcedo et al. 2002; Yatou \& Toh 2009) and
(Mason \& Bessey 1983; Karpen et al. 1989; Trevisan \& Ib\'{a}\~{n}ez 2000), respectively.

In majority of papers studying the thermal instability of astrophysical
objects in the magnetic field, one uses the one-fluid ideal
magnetohydrodynamics (MHD). In some papers, the two-fluid model with the ideal
magnetic induction equation has been treated (e.g., Fukue \& Kamaya 2007;
Inoue \& Inutsuka 2008). The non-ideal effects in the induction equation have
been included in (e.g., Heyvaerts 1974; Stiele et al. 2006; Shadmehri et al. 2010).

In the plasma physics literature devoted to the thermal instability, one
considers besides the one-fluid (e.g., Bora \& Taiwar 1993) and two-fluid
(e.g., Birk 2000) ideal MHD (from the point of view of the form of the
magnetic induction equation) also a multicomponent media where the presence of
electrons, ions, dust grains, and neutrals is taken into account (Kopp et al.
1997; Birk \& Wiechen 2001; Pandey \& Krishan 2001; Pandey et al. 2003; Shukla
\& Sandberg 2003; Kopp \& Shchekinov 2007). Such effect as dust charge
variation is also included when studying the thermal instability
(Ib\'{a}\~{n}ez \& Shchekinov 2002; Pandey \& Krishan 2001; Pandey et al.
2003). Analytical investigation of thermal instability in the multicomponent
magnetized media with such physical effects as collisions between different
species, ionization and recombination, dust charge dynamics, gravity,
self-gravity, rotation, and so on is a very challenging problem. General basic
equations have a very complex form (see, e.g., Kopp et al. 1997). Therefore,
one usually treats simplified models considering, for example, potential
perturbations in nonmagnetized (Kopp et al. 1997; Pandey \& Krishan 2001;
Ib\'{a}\~{n}ez \& Shchekinov 2002; Pandey et al. 2003; Shukla \& Sandberg
2003; Kopp \& Shchekinov 2007) and magnetized (Kopp et al. 1997; Shukla \&
Sandberg 2003) plasmas.

Neglecting the energy exchange between species in the thermal equations is one
of the usual simplifying assumptions. It is justified when a collisional
coupling is weak or when the energy exchange frequency is sufficiently large,
so that the temperatures of species are equal. In a general case, one can
obtain considerable complications. Nevertheless, some authors took into
account the energy exchange for thermal instability in the two-fluid MHD
framework (e.g., Birk 2000; Birk \& Wiechen 2001).

The terms describing the energy exchange contribute to the dispersion relation
not only through the temperature (or pressure) perturbations (Birk 2000; Birk
\& Wiechen 2001) but also through the perturbation of collision frequency
which depends on the density and temperature. This effect is important when
background temperatures of species are different. Such a situation can occur,
for example, in galaxy clusters (see Markevitch et al. 1996; Fox \& Loeb 1997;
Ettori \& Fabian 1998; Takizawa 1998).

In the present paper, we investigate the thermal instability in the
electron-ion magnetized plasma which is relevant in the IGM of galaxy
clusters, solar corona, and other two-component plasma objects. We apply the
multicomponent plasma approach when the dynamics of all species is considered
separately through the electric field perturbations (the $\mathbf{E}%
$-approach; see, e.g., Nekrasov 2009 a, 2009 b, 2009 c; Nekrasov \& Shadmehri
2010, 2011). General expressions obtained in this paper can be applied for a
wide range of astrophysical and laboratory plasmas. We assume that the
background electron and ion temperatures are different and include the energy
exchange in the thermal equations for electrons and ions. We take into account
the dependence of collision frequency on the density and temperature
perturbations. Different cooling-heating functions are assumed for electrons
and ions. We include neither ionization and recombination effects nor gravity.
Some expressions for electron and ion perturbations are obtained in the
general form which can be used for other species (dust grains and neutrals).
Here, we treat the condensation mode of thermal instability in the fast sound
speed limit. We derive the general dispersion relation taking into account the
effects mentioned above and find its simple solutions for the growth rates in
the limiting cases.

The paper is organized in the following manner. In section 2, we give the
fundamental equations used in this paper. The equilibrium state is considered
in section 3. Equations for temperature perturbations are obtained in section
4. In section 5, we give specific conditions for further consideration. In
section 6, equations for components of velocity perturbations are given in the
fast sound limit. Components of the perturbed current are calculated in
section 7. These components for the simplified collision contribution are
given in section 8. In section 9, we derive the dispersion relation. Its
limiting cases are considered in section 10. We discuss the obtained results
in section 11. The possible astrophysical implications are considered in
section 12. Summing up of the main points is given in section 13.

\bigskip

\section{BASIC EQUATIONS}

The fundamental equations that we consider here are the following:%
\begin{equation}
\frac{\partial\mathbf{v}_{j}}{\partial t}+\mathbf{v}_{j}\cdot\mathbf{\nabla
v}_{j}=-\frac{\mathbf{\nabla}p_{j}}{m_{j}n_{j}}+\mathbf{F}_{j}\mathbf{+}%
\frac{q_{j}}{m_{j}c}\mathbf{v}_{j}\times\mathbf{B},
\end{equation}
the equation of motion,%
\begin{equation}
\frac{\partial n_{j}}{\partial t}+\mathbf{\nabla}\cdot n_{j}\mathbf{v}_{j}=0,
\end{equation}
the continuity equation,%
\begin{equation}
\frac{\partial T_{i}}{\partial t}+\mathbf{v}_{i}\cdot\mathbf{\nabla}%
T_{i}+\left(  \gamma-1\right)  T_{i}\mathbf{\nabla}\cdot\mathbf{v}%
_{i}=-\left(  \gamma-1\right)  \frac{1}{n_{i}}%
\mathcal{L}%
_{i}\left(  n_{i},T_{i}\right)  +\nu_{ie}^{\varepsilon}\left(  n_{e}%
,T_{e}\right)  \left(  T_{e}-T_{i}\right)
\end{equation}
and%
\begin{equation}
\frac{\partial T_{e}}{\partial t}+\mathbf{v}_{e}\cdot\mathbf{\nabla}%
T_{e}+\left(  \gamma-1\right)  T_{e}\mathbf{\nabla}\cdot\mathbf{v}%
_{e}=-\left(  \gamma-1\right)  \frac{1}{n_{e}}\mathbf{\nabla\cdot q}%
_{e}-\left(  \gamma-1\right)  \frac{1}{n_{e}}%
\mathcal{L}%
_{e}\left(  n_{e},T_{e}\right)  -\nu_{ei}^{\varepsilon}\left(  n_{i}%
,T_{e}\right)  \left(  T_{e}-T_{i}\right)
\end{equation}
are the temperature equations for ions and electrons. In Equations (1) and
(2), the index $j=i,e$ denotes the ions and electrons, respectively. The force
$\mathbf{F}_{j}$ in Equation (1) is given by%

\begin{align}
\mathbf{F}_{i}  &  =\frac{q_{i}}{m_{i}}\mathbf{E}-\nu_{ie}\left(
\mathbf{v}_{i}-\mathbf{v}_{e}\right)  ,\\
\mathbf{F}_{e}  &  =\frac{q_{e}}{m_{e}}\mathbf{E}-\nu_{ei}\left(
\mathbf{v}_{e}-\mathbf{v}_{i}\right)  .\nonumber
\end{align}
Other notations in Equations (1)-(5) are the following: $q_{j}$ and $m_{j}$
are the charge and mass of species $j=i,e$, $\mathbf{v}_{j}$ is the
hydrodynamic velocity, $n_{j}$ is the number density, $p_{j}=n_{j}T_{j}$ is
the thermal pressure, $T_{j}$ is the temperature, $\nu_{ie}$ ($\nu_{ei}$) is
the collision frequency of ions (electrons) with electrons (ions), $\nu
_{ie}^{\varepsilon}(n_{e},T_{e})=2\nu_{ie}$ ($\nu_{ei}^{\varepsilon}\left(
n_{i},T_{e}\right)  $) is the frequency of the thermal energy exchange between
ions (electrons) and electrons (ions) (Braginskii 1965), $n_{i}\nu
_{ie}^{\varepsilon}\left(  n_{e},T_{e}\right)  =n_{e}\nu_{ei}^{\varepsilon
}\left(  n_{i},T_{e}\right)  $, $\gamma$ is the ratio of specific heats,
$\mathbf{E}$\textbf{\ }and $\mathbf{B}$ are the electric and magnetic fields,
and $c$ is the speed of light in vacuum. The value $\mathbf{q}_{e}$ in
Equation (4) is the electron heat flux (Braginskii 1965). As for the latter,
we will consider a weakly collisional plasma when the electron Larmor radius
is much smaller than the electron collisional mean free path. In this case,
the electron thermal flux is mainly directed along the magnetic field,
\begin{equation}
\mathbf{q}_{e}=-\chi_{e}\mathbf{b}\left(  \mathbf{b\cdot\nabla}\right)  T_{e},
\end{equation}
where $\chi_{e}$ is the electron thermal conductivity coefficient and
$\mathbf{b=B/}B$ is the unit vector along the magnetic field. In other
respects, a relation between cyclotron and collision frequencies of species
stays arbitrary in general expressions considered below. We only take into
account the electron thermal flux (6) because the corresponding ion thermal
conductivity is considerably smaller (Braginskii 1965). We also assume that
the thermal flux in the equilibrium is absent. The cooling and heating of
plasma species in Equations (3) and (4) are described by function $%
\mathcal{L}%
_{j}(n_{j},T_{j})=n_{j}^{2}\Lambda_{j}\left(  T_{j}\right)  -n_{j}\Gamma_{j}$,
where $\Lambda_{j}$ and $\Gamma_{j}$ are the cooling and heating functions,
respectively. The form of this function has some difference from the usually
used cooling-heating function $\pounds $ (Field 1965). Both functions are
connected with each other via equality $%
\mathcal{L}%
_{j}\left(  n_{j},T_{j}\right)  =m_{j}n_{j}\pounds _{j}$. Our choice is
analogous to that as in (Begelman \& Zweibel 1994; Pandey \& Krishan 2001;
Shukla \& Sandberg 2003; Bogdanovi\'{c} et al. 2009; Parrish et al. 2009). The
function $\Lambda_{j}\left(  T_{j}\right)  $ can be found, for example, in
(Tozzi \& Norman 2001).

Electromagnetic equations are Faraday's
\begin{equation}
\mathbf{\nabla\times E=-}\frac{1}{c}\frac{\partial\mathbf{B}}{\partial t}%
\end{equation}
and Ampere`s
\begin{equation}
\mathbf{\nabla\times B=}\frac{4\pi}{c}\mathbf{j}%
\end{equation}
laws, where $\mathbf{j=}\sum_{j}q_{j}n_{j}\mathbf{v}_{j}.$ We consider wave
processes with typical timescales much larger than the time the light spends
to cover the wavelength of perturbations. In this case, one can neglect the
displacement current in Equation (8) that results in quasi-neutrality for both
the electromagnetic and purely electrostatic perturbations. The magnetic field
$\mathbf{B}$ includes the background magnetic field $\mathbf{B}_{0}$, the
magnetic field $\mathbf{B}_{0cur}$ of the background electric current (when it
presents), and the perturbed magnetic field.

For generality, we assume in the meanwhile that $n_{i}$ and $n_{e}$ are
different, having in mind that some solutions obtained below can be applied
for multicomponent plasmas.

\bigskip

\section{EQUILIBRIUM STATE}

At first, we will consider an equilibrium state. We assume that the background
velocities of species are absent. Here, we do not involve an equilibrium
inhomogeneity. Then, the thermal equations (3) and (4) in the equilibrium take
the form%
\begin{align}
\left(  \gamma-1\right)  \frac{1}{n_{i0}}%
\mathcal{L}%
_{i}\left(  n_{i0},T_{i0}\right)  -\nu_{ie}^{\varepsilon}(n_{e0}%
,T_{e0})\left(  T_{e0}-T_{i0}\right)   &  =0,\\
\left(  \gamma-1\right)  \frac{1}{n_{e0}}%
\mathcal{L}%
_{e}\left(  n_{e0},T_{e0}\right)  +\nu_{ei}^{\varepsilon}\left(  n_{i0}%
,T_{e0}\right)  \left(  T_{e0}-T_{i0}\right)   &  =0.\nonumber
\end{align}

\bigskip

\section{LINEAR\ EQUATIONS\ FOR THE TEMPERATURE PERTURBATIONS}

We now consider Equations (3) and (4) in the linear approximations. Using
Equations (2) and (9) for ions, we find%
\begin{equation}
D_{1i}T_{i1}-D_{2i}T_{e1}=C_{1i}\mathbf{\nabla}\cdot\mathbf{v}_{i1}%
-C_{2i}\mathbf{\nabla}\cdot\mathbf{v}_{e1},
\end{equation}
where the index $1$ denotes the perturbed values. The following notations are
here introduced:%
\begin{align}
D_{1i}  &  =\left(  \frac{\partial}{\partial t}+\Omega_{Ti}+\Omega
_{ie}\right)  \frac{\partial}{\partial t},\\
D_{2i}  &  =\left(  \Omega_{Tie}+\Omega_{ie}\right)  \frac{\partial}{\partial
t},\nonumber\\
C_{1i}  &  =T_{i0}\left[  -\left(  \gamma-1\right)  \frac{\partial}{\partial
t}+\Omega_{ni}-\frac{\left(  T_{e0}-T_{i0}\right)  }{T_{i0}}\Omega
_{ie}\right]  ,\nonumber\\
C_{2i}  &  =\Omega_{ie}\left(  T_{e0}-T_{i0}\right)  .\nonumber
\end{align}
where the index $0$ denotes the unperturbed values. Analogously, we obtain for
the electrons%
\begin{equation}
D_{1e}T_{e1}-D_{2e}T_{i1}=C_{1e}\mathbf{\nabla}\cdot\mathbf{v}_{e1}%
+C_{2e}\mathbf{\nabla}\cdot\mathbf{v}_{i1},
\end{equation}
where%
\begin{align}
D_{1e}  &  =\left(  \frac{\partial}{\partial t}+\Omega_{\chi}+\Omega
_{Te}+\Omega_{Tei}+\Omega_{ei}\right)  \frac{\partial}{\partial t},\\
D_{2e}  &  =\Omega_{ei}\frac{\partial}{\partial t},\nonumber\\
C_{1e}  &  =T_{e0}\left[  -\left(  \gamma-1\right)  \frac{\partial}{\partial
t}+\Omega_{ne}+\frac{\left(  T_{e0}-T_{i0}\right)  }{T_{e0}}\Omega
_{ei}\right]  ,\nonumber\\
C_{2e}  &  =\Omega_{ei}\left(  T_{e0}-T_{i0}\right)  .\nonumber
\end{align}
In notations (11) and (13), we have introduced the following frequencies:%
\begin{align}
\Omega_{\chi}  &  =-\left(  \gamma-1\right)  \frac{\chi_{e0}}{n_{e0}}%
\frac{\partial^{2}}{\partial z^{2}},\\
\Omega_{Te}  &  =\left(  \gamma-1\right)  \frac{\partial%
\mathcal{L}%
_{e}\left(  n_{e0},T_{e0}\right)  }{n_{e0}\partial T_{e0}},\Omega_{Ti}=\left(
\gamma-1\right)  \frac{\partial%
\mathcal{L}%
_{i}\left(  n_{i0},T_{i0}\right)  }{n_{i0}\partial T_{i0}},\nonumber\\
\Omega_{ne}  &  =\left(  \gamma-1\right)  \frac{\partial%
\mathcal{L}%
_{e}\left(  n_{e0},T_{e0}\right)  }{T_{e0}\partial n_{e0}},\Omega_{ni}=\left(
\gamma-1\right)  \frac{\partial%
\mathcal{L}%
_{i}\left(  n_{i0},T_{i0}\right)  }{T_{i0}\partial n_{i0}},\nonumber\\
\Omega_{ei}  &  =\nu_{ei}^{\varepsilon}\left(  n_{i0},T_{e0}\right)
,\Omega_{ie}=\nu_{ie}^{\varepsilon}\left(  n_{e0},T_{e0}\right)  ,\nonumber\\
\Omega_{Tei}  &  =\frac{\partial\nu_{ei}^{\varepsilon}\left(  n_{i0}%
,T_{e0}\right)  }{\partial T_{e0}}\left(  T_{e0}-T_{i0}\right)  ,\Omega
_{Tie}=\frac{\partial\nu_{ie}^{\varepsilon}\left(  n_{e0},T_{e0}\right)
}{\partial T_{e0}}\left(  T_{e0}-T_{i0}\right)  .\nonumber
\end{align}
We assume that the background magnetic field $\mathbf{B}_{0}$ is directed
along the $z$-axis. In notations (11) and (13), we have used an equilibrium
state and the fact that $\nu_{ei}^{\varepsilon}\left(  n_{i0},T_{e0}\right)
\sim n_{i0}$ and $\nu_{ie}^{\varepsilon}\left(  n_{e0},T_{e0}\right)  \sim
n_{e0}$. We see from Equations (10) and (12) that the temperature
perturbations are connected with a velocity divergence. Solutions for $T_{e1}
$ and $T_{i1}$ are given by%
\begin{equation}
DT_{e1}=G_{1}\ \mathbf{\nabla}\cdot\mathbf{v}_{e1}+G_{2}\mathbf{\nabla}%
\cdot\mathbf{v}_{i1},
\end{equation}%
\begin{equation}
DT_{i1}=G_{3}\mathbf{\nabla}\cdot\mathbf{v}_{e1}+G_{4}\mathbf{\nabla}%
\cdot\mathbf{v}_{i1},
\end{equation}
where the following notations are introduced:%
\begin{align}
D  &  =D_{1i}D_{1e}-D_{2i}D_{2e},\\
G_{1}  &  =D_{1i}C_{1e}-D_{2e}C_{2i},\nonumber\\
G_{2}  &  =D_{1i}C_{2e}+D_{2e}C_{1i},\nonumber\\
G_{3}  &  =D_{2i}C_{1e}-D_{1e}C_{2i},\nonumber\\
G_{4}  &  =D_{1e}C_{1i}+D_{2i}C_{2e}.\nonumber
\end{align}

To find the temperature perturbation $T_{j1}$, we have to calculate
expressions for $\mathbf{\nabla}\cdot\mathbf{v}_{j1}$. General equations for
the velocity $\mathbf{v}_{j1}$ and $\mathbf{\nabla}\cdot\mathbf{v}_{j1} $ are
derived in the Appendix, where expressions for $D$ and $G_{l}$, $l=1,2,3,4$,
are also given. In their general form, the components of $\mathbf{v}_{j1}$ are
very complex. Therefore to proceed further analytically, here we restrict
ourselves to a limiting case in which the dynamical frequency $\partial
/\partial t$ is smaller than the sound frequency. Thus, we will consider
sufficiently short-wavelength perturbations along the magnetic field (see
below). Some additional simplifying conditions which are satisfied in
magnetized plasmas are also used.

\bigskip

\section{SPECIFIC CASE: FAST SOUND SPEED LIMIT}

Equations (A27), (A28) and (A30) are written in their general form which
allows us to consider different simplified specific cases corresponding to the
real astrophysical conditions. We further consider the case in which%
\begin{equation}
\omega_{ci}^{2}\gg\left(  \frac{\partial}{\partial z}\right)  ^{-2}\left(
\frac{\partial^{2}}{\partial y^{2}}+\frac{\partial^{2}}{\partial z^{2}%
}\right)  \frac{\partial^{2}}{\partial t^{2}}.
\end{equation}
Since usually $\omega_{ci}^{2}\gg\partial^{2}/\partial t^{2}$ in magnetized
plasmas, condition (18) is satisfied for a wide range of the transverse
wavelengths of perturbations $\left(  \partial/\partial y\right)  ^{-1}$ in
comparison with the longitudinal wavelengths $\left(  \partial/\partial
z\right)  ^{-1}$. The other simplifying condition is\textbf{\ }
\begin{equation}
\omega_{ci}^{2}\gg\frac{\left[  T_{e0}T_{i0}\frac{\partial}{\partial
t}+\left(  T_{e0}+T_{i0}\right)  ^{2}\Omega_{ie}\right]  }{m_{i}\left[
T_{e0}\frac{\partial}{\partial t}+\left(  T_{e0}+T_{i0}\right)  \Omega
_{ie}\right]  }\left(  \frac{\partial^{2}}{\partial y^{2}}+\frac{\partial^{2}%
}{\partial z^{2}}\right)  .
\end{equation}
This inequality means that the effective ion Larmor radius is much smaller than the
perturbations wavelengths.

Below, we consider the case in which the perturbation frequency is much
smaller than the sound frequency, i.e.,
\begin{equation}
\frac{\partial^{2}}{\partial t^{2}}\ll\frac{\left[  T_{e0}T_{i0}\frac
{\partial}{\partial t}+\left(  T_{e0}+T_{i0}\right)  ^{2}\Omega_{ie}\right]
}{m_{i}\left[  T_{e0}\frac{\partial}{\partial t}+\left(  T_{e0}+T_{i0}\right)
\Omega_{ie}\right]  }\frac{\partial^{2}}{\partial z^{2}}.
\end{equation}
This condition can be written in the form $\partial^{2}/\partial t^{2}\ll
c_{s}^{2}\partial^{2}/\partial z^{2}$, where $c_{s}$ can be considered as the
effective sound velocity. Inequalities (19) and (20) are given in the
approximate form to unite two cases, $\frac{\partial}{\partial t}%
>(<)\Omega_{ie,ei}$. We note that the case (18) is followed from conditions
(19) and (20).

Under conditions (19) and (20), we find, using notations (A10), (A30), and
(A31), the following equations for $P_{i,e1}$ (see Equations (A27) and (A28)):%
\begin{equation}
\frac{\partial^{2}P_{i1}}{\partial z^{2}}=-\left(  1-\alpha_{i}\right)
\left(  \frac{1}{\omega_{ci}^{2}}\frac{\partial^{3}Q_{i1y}}{\partial y\partial
t^{2}}\ +\frac{\partial^{2}F_{i1z}}{\partial z\partial t}\right)  +\beta
_{e}\left(  \frac{1}{\omega_{ce}^{2}}\frac{\partial^{3}Q_{e1y}}{\partial
y\partial t^{2}}+\frac{\partial^{2}F_{e1z}}{\partial z\partial t}\right)  ,
\end{equation}%
\begin{equation}
\frac{\partial^{2}P_{e1}}{\partial z^{2}}=-\left(  1-\alpha_{e}\right)
\left(  \frac{1}{\omega_{ce}^{2}}\frac{\partial^{3}Q_{e1y}}{\partial y\partial
t^{2}}+\frac{\partial^{2}F_{e1z}}{\partial z\partial t}\right)  +\beta
_{i}\left(  \frac{1}{\omega_{ci}^{2}}\frac{\partial^{3}Q_{i1y}}{\partial
y\partial t^{2}}+\frac{\partial^{2}F_{i1z}}{\partial z\partial t}\right)  ,
\end{equation}
where%
\begin{align}
\alpha_{i} &  =\frac{m_{i}}{K}\left(  DT_{e0}-G_{1}\frac{\partial}{\partial
t}\right)  \left(  \frac{\partial}{\partial z}\right)  ^{-2}\frac{\partial
^{2}}{\partial t^{2}},\\
\alpha_{e} &  =\frac{m_{e}}{K}\left(  DT_{i0}-G_{4}\frac{\partial}{\partial
t}\right)  \left(  \frac{\partial}{\partial z}\right)  ^{-2}\frac{\partial
^{2}}{\partial t^{2}},\nonumber\\
\beta_{i} &  =\frac{m_{i}}{K}G_{2}\left(  \frac{\partial}{\partial z}\right)
^{-2}\frac{\partial^{3}}{\partial t^{3}},\nonumber\\
\beta_{e} &  =\frac{m_{e}}{K}G_{3}\left(  \frac{\partial}{\partial z}\right)
^{-2}\frac{\partial^{3}}{\partial t^{3}}.\nonumber
\end{align}

Below, we will need the correction terms proportional to $\alpha_{e,i}\ll1$
and $\beta_{e,i}\ll1$. We note that the contribution of the correction term
proportional to $Q_{e1y}$ in Equation (21) is at least of the order of
$m_{e}/m_{i}$ in comparison with the term proportional to $Q_{i1y}$. However
for convenience, we keep this small term for the symmetry with equation for
$P_{e1}$. The ratio of contributions of the correction terms proportional to
$F_{e1z}$ and $F_{i1z}$ in Equation (21) is of the order of
\[
\frac{\left(  T_{e0}+T_{i0}\right)  \Omega_{ie}}{T_{e0}\frac{\partial
}{\partial t}+\left(  T_{e0}+T_{i0}\right)  \Omega_{ie}}.
\]
The same for terms proportional to $Q_{i1y}$ and $Q_{e1y}$ in Equation (22)
is
\[
\frac{m_{i}}{m_{e}}\frac{\left(  T_{e0}+T_{i0}\right)  \Omega_{ei}}%
{T_{i0}\frac{\partial}{\partial t}+\left(  T_{e0}+T_{i0}\right)  \Omega_{ei}},
\]
and for the ratio of terms $\sim F_{i1z}$ and $F_{e1z}$, we have%
\[
\frac{\left(  T_{e0}+T_{i0}\right)  \Omega_{ei}}{T_{i0}\frac{\partial
}{\partial t}+\left(  T_{e0}+T_{i0}\right)  \Omega_{ei}}.
\]

\section{EQUATIONS FOR COMPONENTS OF VELOCITIES $\mathbf{v}_{i,e1}$}

We now obtain equations for components of velocities $\mathbf{v}_{i,e1}$,
using Equations (21) and (22).

\subsection{Equations for $v_{i,e1y}$}

From Equations (A3), (21), and (22), we find, using notations (A6),%
\begin{align}
v_{i1y}  &  =-\frac{1}{\omega_{ci}}F_{i1x}+\frac{1}{\omega_{ci}^{3}}\left[
1+\frac{\partial^{2}}{\partial y^{2}}\left(  \frac{\partial}{\partial
z}\right)  ^{-2}\right]  \frac{\partial^{2}F_{i1x}}{\partial t^{2}}+\frac
{1}{\omega_{ci}^{2}}\frac{\partial F_{i1y}}{\partial t}\\
&  -\frac{1}{\omega_{ci}^{4}}\left[  1+\frac{\partial^{2}}{\partial y^{2}%
}\left(  \frac{\partial}{\partial z}\right)  ^{-2}\right]  \frac{\partial
^{3}F_{i1y}}{\partial t^{3}}-\beta_{e}\frac{1}{\omega_{ce}\omega_{ci}^{2}%
}\left(  \frac{\partial}{\partial z}\right)  ^{-2}\frac{\partial^{4}F_{e1x}%
}{\partial y^{2}\partial t^{2}}\nonumber\\
&  +\beta_{e}\frac{1}{\omega_{ce}^{2}\omega_{ci}^{2}}\left(  \frac{\partial
}{\partial z}\right)  ^{-2}\frac{\partial^{5}F_{e1y}}{\partial y^{2}\partial
t^{3}}-\frac{1}{\omega_{ci}^{2}}\left(  \frac{\partial}{\partial z}\right)
^{-1}\frac{\partial^{2}F_{i1z}}{\partial y\partial t}\nonumber\\
&  +\frac{1}{\omega_{ci}^{2}}\left(  \frac{\partial}{\partial z}\right)
^{-1}\frac{\partial^{2}}{\partial y\partial t}\left(  \alpha_{i}F_{i1z}%
+\beta_{e}F_{e1z}\right)  ,\nonumber
\end{align}%
\begin{align}
v_{e1y}  &  =-\frac{1}{\omega_{ce}}F_{e1x}+\frac{1}{\omega_{ce}^{3}}\left[
1+\ \frac{\partial^{2}}{\partial y^{2}}\left(  \frac{\partial}{\partial
z}\right)  ^{-2}\right]  \frac{\partial^{2}F_{e1x}}{\partial t^{2}}+\frac
{1}{\omega_{ce}^{2}}\frac{\partial F_{e1y}}{\partial t}\\
&  -\frac{1}{\omega_{ce}^{4}}\left[  1+\frac{\partial^{2}}{\partial y^{2}%
}\left(  \frac{\partial}{\partial z}\right)  ^{-2}\right]  \frac{\partial
^{3}F_{e1y}}{\partial t^{3}}-\beta_{i}\frac{1}{\omega_{ci}\omega_{ce}^{2}%
}\ \left(  \frac{\partial}{\partial z}\right)  ^{-2}\frac{\partial^{4}F_{i1x}%
}{\partial y^{2}\partial t^{2}}\nonumber\\
&  +\beta_{i}\frac{1}{\omega_{ci}^{2}\omega_{ce}^{2}}\ \left(  \frac{\partial
}{\partial z}\right)  ^{-2}\frac{\partial^{5}F_{i1y}}{\partial y^{2}\partial
t^{3}}-\frac{1}{\omega_{ce}^{2}}\ \left(  \frac{\partial}{\partial z}\right)
^{-1}\frac{\partial^{2}F_{e1z}}{\partial y\partial t}\nonumber\\
&  +\frac{1}{\omega_{ce}^{2}}\ \left(  \frac{\partial}{\partial z}\right)
^{-1}\frac{\partial^{2}}{\partial y\partial t}\left(  \alpha_{e}F_{e1z}%
+\beta_{i}F_{i1z}\right)  .\nonumber
\end{align}
We see that these equations are obtained one from another by changing
$i\leftrightarrow e$. We note that $\alpha_{j},\beta_{j}\sim m_{j}$. However,
we keep in these equations some small terms for the symmetry of both
equations. The terms proportional to $\omega_{cj}^{-4}$ are needed in
equations for $v_{i,e1x}$. We also note that $\alpha_{i}\sim\left(
m_{i}/T_{0}\right)  \left(  \partial/\partial z\right)  ^{-2}\partial
^{2}/\partial t^{2}$ and $\beta_{i}\sim\alpha_{i}\Omega_{ei}/\left(
\partial/\partial t+\Omega_{ei}\right)  $, if $T_{e0}\sim T_{i0}\sim T_{0}$.
For these estimations, we have used expressions (A18), (A25), (A26), and
(A29), assuming $W_{j}\sim\partial/\partial t$.

\subsection{Equations for $v_{i,e1x}$}

Equations for $v_{i,e1x}$ are found from Equation (A2) by using Equations (24)
and (25).

\subsection{Equations for $v_{i,e1z}$}

From Equations (A7), (21), and (22), we obtain, using notations (A6),%
\begin{align}
\frac{\partial^{2}v_{i1z}}{\partial z\partial t}  &  =\frac{1}{\omega_{ci}%
}\frac{\partial^{2}F_{i1x}}{\partial y\partial t}-\frac{\partial^{2}}{\partial
y\partial t}\left(  \frac{\alpha_{i}}{\omega_{ci}}F_{i1x}+\frac{\beta_{e}%
}{\omega_{ce}}F_{e1x}\right) \\
&  -\frac{1}{\omega_{ci}^{2}}\frac{\partial^{3}F_{i1y}}{\partial y\partial
t^{2}}+\frac{\partial^{3}}{\partial y\partial t^{2}}\left(  \frac{\alpha_{i}%
}{\omega_{ci}^{2}}F_{i1y}+\frac{\beta_{e}}{\omega_{ce}^{2}}F_{e1y}\right)
\nonumber\\
&  +\frac{\partial}{\partial z}\left(  \alpha_{i}F_{i1z}+\beta_{e}%
F_{e1z}\right)  ,\nonumber
\end{align}%
\begin{align}
\frac{\partial^{2}v_{e1z}}{\partial z\partial t}  &  =\frac{1}{\omega_{ce}%
}\frac{\partial^{2}F_{e1x}}{\partial y\partial t}-\frac{\partial^{2}}{\partial
y\partial t}\left(  \frac{\alpha_{e}}{\omega_{ce}}F_{e1x}+\frac{\beta_{i}%
}{\omega_{ci}}F_{i1x}\right) \\
&  -\frac{1}{\omega_{ce}^{2}}\frac{\partial^{3}F_{e1y}}{\partial y\partial
t^{2}}+\frac{\partial^{3}}{\partial y\partial t^{2}}\left(  \frac{\alpha_{e}%
}{\omega_{ce}^{2}}F_{e1y}+\frac{\beta_{i}}{\omega_{ci}^{2}}F_{i1y}\right)
\nonumber\\
&  +\frac{\partial}{\partial z}\left(  \alpha_{e}F_{e1z}+\beta_{i}%
F_{i1z}\right)  .\nonumber
\end{align}
We note that in Equation (26) ((27)) the terms proportional to $\beta
_{e}F_{e1x,y}$ ($\alpha_{e}F_{e1x,y}$) are small as compared with $\alpha
_{i}F_{i1x,y}$ ($\beta_{i}F_{i1x,y}$). However, we also keep them for the
symmetry of these equations.

\bigskip

\section{COMPONENTS OF CURRENT}

We now find components of the linear current $\mathbf{j}_{1}=%
{\displaystyle\sum\limits_{j}}
q_{j}n_{j0}\mathbf{v}_{j1}$. It is convenient to calculate the value
$4\pi\left(  \partial/\partial t\right)  ^{-1}\mathbf{j}_{1}$. We further
consider the electron-ion plasma in which $n_{e0}=n_{i0}$, $q_{e}=-q_{i}$. In
our calculations, we will use an equality $m_{e}\nu_{ei}=m_{i}\nu_{ie}$. From
Equations (A2), (24)-(27), and (5) in the linear approximation, we find
\begin{align}
4\pi\left(  \frac{\partial}{\partial t}\right)  ^{-1}j_{1x} &  =a_{xx}%
E_{1x}-a_{xy}E_{1y}+a_{xz}E_{1z}\\
&  -b_{xx}\left(  v_{i1x}-v_{e1x}\right)  +b_{xy}\left(  v_{i1y}%
-v_{e1y}\right)  -b_{xz}\left(  v_{i1z}-v_{e1z}\right)  ,\nonumber
\end{align}%
\begin{align}
4\pi\left(  \frac{\partial}{\partial t}\right)  ^{-1}j_{1y} &  =a_{yx}%
E_{1x}+a_{yy}E_{1y}-a_{yz}E_{1z}\\
&  -b_{yx}\left(  v_{i1x}-v_{e1x}\right)  -b_{yy}\left(  v_{i1y}%
-v_{e1y}\right)  +b_{yz}\left(  v_{i1z}-v_{e1z}\right)  ,\nonumber
\end{align}%
\begin{align}
4\pi\left(  \frac{\partial}{\partial t}\right)  ^{-1}j_{1z} &  =-a_{zx}%
E_{1x}-a_{zy}E_{1y}+a_{zz}E_{1z}\\
&  +b_{zx}\left(  v_{i1x}-v_{e1x}\right)  +b_{zy}\left(  v_{i1y}%
-v_{e1y}\right)  -b_{zz}\left(  v_{i1z}-v_{e1z}\right)  .\nonumber
\end{align}
Here, the following notations are introduced: \
\begin{align}
a_{xx} &  =\frac{\omega_{pi}^{2}}{\omega_{ci}^{2}}\left[  1+\frac{\partial
^{2}}{\partial y^{2}}\left(  \frac{\partial}{\partial z}\right)  ^{-2}\right]
,a_{xy}=a_{yx}=\frac{\omega_{pi}^{2}}{\omega_{ci}^{3}}\left[  1+\frac
{\partial^{2}}{\partial y^{2}}\left(  \frac{\partial}{\partial z}\right)
^{-2}\right]  \frac{\partial}{\partial t},\\
a_{xz} &  =\frac{\omega_{pi}^{2}}{\omega_{ci}}\left(  \alpha_{i}-\beta
_{e}\frac{m_{i}}{m_{e}}\right)  \frac{\partial}{\partial y}\left(
\frac{\partial^{2}}{\partial z\partial t}\right)  ^{-1},a_{yy}=\frac
{\omega_{pi}^{2}}{\omega_{ci}^{2}},a_{yz}=a_{zy}=\frac{\omega_{pi}^{2}}%
{\omega_{ci}^{2}}\frac{\partial}{\partial y}\left(  \frac{\partial}{\partial
z}\right)  ^{-1},\nonumber\\
a_{zx} &  =\frac{\omega_{pi}^{2}}{\omega_{ci}}\left(  \alpha_{i}-\beta
_{i}\right)  \frac{\partial}{\partial y}\left(  \frac{\partial^{2}}{\partial
z\partial t}\right)  ^{-1},a_{zz}=\omega_{pi}^{2}m_{i}\delta\left(
\frac{\partial}{\partial t}\right)  ^{-2},b_{ij}=a_{ij}\frac{m_{i}}{q_{i}}%
\nu_{ie},\nonumber
\end{align}
where $\omega_{pi}=\left(  4\pi n_{i0}q_{i}^{2}/m_{i}\right)  ^{1/2}$ is the
ion plasma frequency. The value $\delta$ in expression for $a_{zz}$ has the
form
\begin{equation}
\delta=\frac{\left(  \alpha_{i}-\beta_{i}\right)  }{m_{i}}+\frac{\left(
\alpha_{e}-\beta_{e}\right)  }{m_{e}}.
\end{equation}

\bigskip

\section{SIMPLIFICATION\ OF\ COLLISION\ CONTRIBUTION}

The relationship between $\omega_{ce}$ and $\nu_{ei}$ or $\omega_{ci}$ and
$\nu_{ie}$ (that is the same) can be arbitrary in Equations (28)-(30) (except
of that in the thermal conduction). We further proceed by taking into account
that $\partial/\partial t\ll\omega_{ci}$. In this case, we can neglect the
collisional terms proportional to $b_{xy}$ and $b_{yx}$ (see notations (31)).
However, a system of Equations (28)-(30) remains sufficiently complex to find
$\mathbf{j}_{1}$ through $\mathbf{E}_{1}$. Therefore, we further consider the
case in which the following condition is satisfied:%

\begin{equation}
1\gg\frac{\nu_{ie}}{\omega_{ci}^{2}}\frac{\partial}{\partial t}\left(
\frac{\partial^{2}}{\partial y^{2}}+\frac{\partial^{2}}{\partial z^{2}%
}\right)  \left(  \frac{\partial}{\partial z}\right)  ^{-2}.
\end{equation}
It is clear that condition (33) can easily be realized. In this case, we may
neglect collisional terms proportional to $b_{x,zx}$ and $b_{y,zy}$. Then, the
system of Equations (28)-(30) takes the form%
\begin{equation}
4\pi\left(  \frac{\partial}{\partial t}\right)  ^{-1}j_{1x}=\varepsilon
_{xx}E_{1x}-\varepsilon_{xy}E_{1y}+\varepsilon_{xz}E_{1z},
\end{equation}%
\begin{equation}
4\pi\left(  \frac{\partial}{\partial t}\right)  ^{-1}j_{1y}=\varepsilon
_{yx}E_{1x}+\varepsilon_{yy}E_{1y}-\varepsilon_{yz}E_{1z},
\end{equation}%
\begin{equation}
4\pi\left(  \frac{\partial}{\partial t}\right)  ^{-1}j_{1z}=-\varepsilon
_{zx}E_{1x}-\varepsilon_{zy}E_{1y}+\varepsilon_{zz}E_{1z}.
\end{equation}
Here,
\begin{align}
\varepsilon_{xx} &  =a_{xx},\varepsilon_{xy}=\left[  a_{xy}-\frac{\nu_{ie}%
}{\omega_{ci}^{2}}\frac{a_{xz}}{\left(  1+d\right)  }\frac{\partial^{2}%
}{\partial y\partial t}\left(  \frac{\partial}{\partial z}\right)
^{-1}\right]  ,\varepsilon_{xz}=\frac{a_{xz}}{\left(  1+d\right)  },\\
\varepsilon_{yx} &  =\left[  a_{yx}-\frac{\nu_{ie}}{\omega_{ci}^{2}}%
\frac{a_{zx}}{\left(  1+d\right)  }\frac{\partial^{2}}{\partial y\partial
t}\left(  \frac{\partial}{\partial z}\right)  ^{-1}\right]  ,\varepsilon
_{yy}=a_{yy},\varepsilon_{yz}=\frac{a_{yz}}{\left(  1+d\right)  },\nonumber\\
\varepsilon_{zx} &  =\frac{a_{zx}}{\left(  1+d\right)  },\varepsilon
_{zy}=\frac{a_{zy}}{\left(  1+d\right)  },\varepsilon_{zz}=\frac{a_{zz}%
}{\left(  1+d\right)  },\nonumber
\end{align}
where
\begin{equation}
d=a_{zz}\frac{\nu_{ie}}{\omega_{pi}^{2}}\frac{\partial}{\partial t}=\nu
_{ie}m_{i}\delta\left(  \frac{\partial}{\partial t}\right)  ^{-1}.
\end{equation}
Parameter $d$ defines the collisionless, $d\ll1$, and collisional, $d\gg1$,
regimes. Below, we will derive the dispersion relation.

\bigskip

\section{DISPERSION\ RELATION AND\ ELECTRIC\ FIELD POLARIZATION\ }

We will further consider Equations (34)-(36) in the Fourier-representation,
assuming that perturbations have the form $\exp\left(  i\mathbf{kr-}i\omega
t\right)  $. Then using Equations (7) and (8), we obtain the following system
of equations:%
\begin{align}
\left(  n^{2}-\varepsilon_{xx}\right)  E_{1xk}+\varepsilon_{xy}E_{1yk}%
-\varepsilon_{xz}E_{1zk}  &  =0,\\
-\varepsilon_{yx}E_{1xk}+\left(  n_{z}^{2}-\varepsilon_{yy}\right)
E_{1yk}+\left(  -n_{y}n_{z}+\varepsilon_{yz}\right)  E_{1zk}  &
=0,\nonumber\\
\varepsilon_{zx}E_{1xk}+\left(  -n_{y}n_{z}+\varepsilon_{zy}\right)
E_{1yk}+\left(  n_{y}^{2}-\varepsilon_{zz}\right)  E_{1zk}  &  =0,\nonumber
\end{align}
where $\mathbf{E}_{1k}$ is the Fourier-image of the electric field
perturbation, $\mathbf{n=k}c/\omega$. The index $k$ by $\mathbf{E}_{1k}$ is
equal to $k=\left\{  \mathbf{k,}\omega\right\}  $. For the Fourier-images of
operators $\varepsilon_{ij}$ and $d$, we keep the same notations. In general,
we see that the longitudinal electric field $E_{1zk}\sim E_{1x,yk}$ inevitably
arises when $k_{y}\neq0$ and $n_{y}^{2}-\varepsilon_{zz}\neq0$. The dispersion
relation can be found by setting the determinant of the system (39) equal to
zero. Contribution of terms proportional to $\varepsilon_{xy}$ and
$\varepsilon_{yx}$ into dispersion relation is small as compared with that of
terms $\sim\varepsilon_{xx},\varepsilon_{yy}$ according to condition (18) and
$\alpha_{i},\beta_{i}\ll1$. Neglecting these small terms, we obtain \
\begin{align}
\left(  n^{2}-\varepsilon_{xx}\right)  E_{1xk}-\varepsilon_{xz}E_{1zk}  &
=0,\\
\left(  n_{z}^{2}-\varepsilon_{yy}\right)  E_{1yk}+\left(  -n_{y}%
n_{z}+\varepsilon_{yz}\right)  E_{1zk}  &  =0,\nonumber\\
\varepsilon_{zx}E_{1xk}+\left(  -n_{y}n_{z}+\varepsilon_{zy}\right)
E_{1yk}+\left(  n_{y}^{2}-\varepsilon_{zz}\right)  E_{1zk}  &  =0.\nonumber
\end{align}
Below, we will consider the dispersion relation for this system in the
collisionless $\left(  d\ll1\right)  $ and collisional $\left(  d\gg1\right)
$ cases.

\subsection{Collisionless case}

Consider the case%
\begin{equation}
d\ll1,
\end{equation}
where an estimation of $d$ is $d\sim\nu_{ie}\omega/k_{z}^{2}c_{s}^{2}$ for
$\delta\sim\alpha_{i}/m_{i}$. Then the dispersion relation has two solutions
when $n_{z}^{2}-a_{yy}=0$ and $n_{z}^{2}-a_{yy}\neq0$ (see the second equation
in the system (40)).

\paragraph{The case $n_{z}^{2}-a_{yy}=0$}

This case describes the Alfv\'{e}n waves $\omega^{2}=k_{z}^{2}c_{A}^{2}$,
where $c_{A}=B_{0}/\left(  4\pi n_{i}m_{i}\right)  ^{1/2}$. Polarization of
the electric field has the form $E_{1yk}\neq0$ and $E_{1xk}=E_{1zk}=0$.

\paragraph{The case $n_{z}^{2}-a_{yy}\neq0$}

In this case, the dispersion relation reduces to $\varepsilon_{zz}\simeq0$.
Contribution of all other terms can be shown, using expressions (31) and (37),
to be small. The last equation means that
\begin{equation}
\delta\simeq0.
\end{equation}
Polarization of these perturbations is the following:%
\begin{align}
E_{1zk} &  \neq0,\\
E_{1yk} &  =\frac{n_{y}}{n_{z}}E_{1zk},\nonumber\\
E_{1xk} &  =\frac{a_{xz}}{\left(  n^{2}-a_{xx}\right)  }E_{1zk}.\nonumber
\end{align}
We see that the electric field in the plane of the wave vector is a potential
one. However, the component $E_{1xk}$ can be large, $E_{1xk}\sim\alpha
_{i}\left(  k_{y}k_{z}\omega_{ci}/k^{2}\omega\right)  E_{1zk}$. Thus, in
general, perturbations have an electromagnetic nature.

\subsection{Collisional case}

In the collisional case,
\begin{equation}
d\gg1,
\end{equation}
the dispersion relation of the system (40) takes the form%
\begin{equation}
\left(  n_{z}^{2}-a_{yy}\right)  a_{zz}=0.
\end{equation}
We see from Equation (45) that the same cases, $n_{z}^{2}-a_{yy}=0$ and
$a_{zz}=0$, are possible analogously to the case $d\ll1$ considered above.
However, polarization in the case $n_{z}^{2}-a_{yy}=0$ is different:
$E_{1zk}=0,$ $E_{1yk}\neq0,$ and $E_{1xk}=\left(  n_{y}n_{z}d/a_{zx}\right)
E_{1yk}\sim\left(  \nu_{ie}/\omega_{ci}\right)  E_{1yk}$. This is a mixture of
the Alfv\'{e}n and magnetosonic waves. For the dispersion relation $a_{zz}=0$,
the electric field polarization is the following:%
\begin{align}
E_{1zk} &  \neq0,\\
E_{1xk} &  =\frac{a_{xz}}{\left(  n^{2}-a_{xx}\right)  d}E_{1zk},\nonumber\\
E_{1yk} &  =\frac{\left(  n_{y}n_{z}d-a_{yz}\right)  }{\left(  n_{z}%
^{2}-a_{yy}\right)  d}E_{1zk},\nonumber
\end{align}
where an estimation for $d$ is given above. Thus, the electric field
perturbation has all three components.

We see that in the case $\varepsilon_{zz}=0$ the electric field perturbation
along the magnetic field plays the crucial role in the fast sound speed regime (20).

\bigskip

\section{SOLUTION\ OF\ DISPERSION\ RELATION\ $\varepsilon_{zz}=0$}

We now consider the dispersion relation (42) which is suitable also for the
collisional case. Using notations (23), (32), and (A16)-(A20), we obtain the
following dispersion relation:%

\begin{align}
&  T_{e0}\left(  \gamma\frac{\partial}{\partial t}+\Omega_{\chi}+\Omega
_{Te}-\Omega_{ne}\right)  \left(  \frac{\partial}{\partial t}+\Omega
_{Ti}+2\Omega_{ie}\right) \\
&  +T_{i0}\left(  \gamma\frac{\partial}{\partial t}+\Omega_{Ti}-\Omega
_{ni}\right)  \left(  \frac{\partial}{\partial t}+\Omega_{\chi}+\Omega
_{Te}+2\Omega_{ei}\right) \nonumber\\
&  -\Omega_{ie}\left(  T_{i0}-T_{e0}\right)  \left(  \frac{\partial}{\partial
t}+\Omega_{\chi}+\Omega_{Te}\right)  -\Omega_{ei}\left(  T_{e0}-T_{i0}\right)
\left(  \frac{\partial}{\partial t}+\Omega_{Ti}\right) \nonumber\\
&  +T_{e0}\Omega_{Tei}\left(  \frac{\partial}{\partial t}+\Omega_{Ti}\right)
+T_{i0}\Omega_{Tei}\left(  \gamma\frac{\partial}{\partial t}+\Omega
_{Ti}-\Omega_{ni}\right)  -T_{e0}\Omega_{Tie}\left[  \Omega_{ne}-\left(
\gamma-1\right)  \frac{\partial}{\partial t}\right]  =0,\nonumber
\end{align}
where $\partial/\partial t=-i\omega$ and $\Omega_{\chi}=\left(  \gamma
-1\right)  \left(  \chi_{e0}/n_{e0}\right)  k_{z}^{2}$. All terms $\Omega$ are
defined by the system (14). We see that the first four terms on the left
hand-side of Equation (47) are symmetric concerning the contribution of
electrons and ions. The third and fourth terms have appeared due to dependence
of collision frequency $\nu_{ei,ie}^{\varepsilon}\left(  n_{i,e0}%
,T_{e}\right)  $ on density perturbation. The last three terms are connected
with perturbation of this frequency because of the electron temperature
perturbation. All terms proportional to $\Omega_{ei,ie}$ and $\Omega_{Tei,ie}$
are connected with the energy exchange in the thermal equations (3) and (4).
We see that taking into account the collision frequency perturbation in a
general case $T_{e0}\neq T_{i0}$ results in considerable modification of the
dispersion relation. However, this effect should be involved because the
absence of the thermal equilibrium between electrons and ions can occur, for
example, in galaxy clusters (e.g., Markevitch et al. 1996; Fox \& Loeb 1997;
Ettori \& Fabian 1998; Takizawa 1998). We further use the fact that
$n_{i0}=n_{e0}$ in our case and $\nu_{ei,ie}^{\varepsilon}\sim T_{e}^{-3/2}$
(Braginskii 1965). Then, we obtain $\Omega_{ei}=\Omega_{ie}$ and $\Omega
_{Tei}=\Omega_{Tie}=-3\Omega_{ie}\left(  T_{e0}-T_{i0}\right)  /2T_{e0}$. In
this case, Equation (47) takes the form%
\begin{align}
&  T_{e0}\left(  \gamma\frac{\partial}{\partial t}+\Omega_{\chi}+\Omega
_{Te}-\Omega_{ne}\right)  \left(  \frac{\partial}{\partial t}+\Omega
_{Ti}+2\Omega_{ie}\right) \\
&  +T_{i0}\left(  \gamma\frac{\partial}{\partial t}+\Omega_{Ti}-\Omega
_{ni}\right)  \left(  \frac{\partial}{\partial t}+\Omega_{\chi}+\Omega
_{Te}+2\Omega_{ie}\right) \nonumber\\
&  +\Omega_{ie}\left(  T_{e0}-T_{i0}\right)  \left(  \Omega_{\chi}+\Omega
_{Te}-\Omega_{Ti}\right) \nonumber\\
&  -\frac{3}{2}\Omega_{ie}\left(  T_{e0}-T_{i0}\right)  \left[  \left(
\gamma\frac{\partial}{\partial t}+\Omega_{Ti}\right)  \left(  1+\frac{T_{i0}%
}{T_{e0}}\right)  -\Omega_{ne}-\frac{T_{i0}}{T_{e0}}\Omega_{ni}\right]
=0.\nonumber
\end{align}
Below, we will consider different limiting cases of Equation (48).

\subsection{The case $\Omega_{ie}=0$}

If we do not take into account the energy exchange, $\Omega_{ie}=0$, and set
$T_{e0}=T_{i0}$, then we obtain equation%

\begin{align}
&  2\gamma\omega^{2}+i\left[  \left(  \gamma+1\right)  \left(  \Omega_{\chi
}+\Omega_{Te}+\Omega_{Ti}\right)  -\Omega_{ne}-\Omega_{ni}\right]  \omega\\
&  -2\left(  \Omega_{\chi}+\Omega_{Te}\right)  \Omega_{Ti}+\Omega_{ne}%
\Omega_{Ti}+\Omega_{ni}\left(  \Omega_{\chi}+\Omega_{Te}\right)  =0.\nonumber
\end{align}
It is followed from Equation (49) that the ion cooling-heating function
modifies the growth rate which takes place without this function. Neglecting
the contribution of the ion cooling and heating, $\Omega_{Ti}=\Omega_{ni}=0 $,
we have%
\[
\omega=-\frac{i}{2\gamma}\left[  \left(  \gamma+1\right)  \left(  \Omega
_{\chi}+\Omega_{Te}\right)  -\Omega_{ne}\right]  .
\]
It is easy to see that this solution is a mixture of isochoric and isobaric
solutions (Parker 1953; Field 1965) because we have taken into account the ion
temperature perturbation. If we neglect the latter, i.e. neglect the second
term $\sim T_{i0}$ in Equation (48), we obtain the usual isobaric solution
\[
\omega=-\frac{i}{\gamma}\left(  \Omega_{\chi}+\Omega_{Te}-\Omega_{ne}\right)
.
\]
We also see from Equation (48) that for the short-wavelength perturbations
when $\Omega_{\chi}\gg\omega,\Omega_{Te},\Omega_{ne}$ the thermal instability
can arise due to the ion cooling function%
\[
\omega=-\frac{i}{T_{e0}+\gamma T_{i0}}\left[  \left(  T_{e0}+T_{i0}\right)
\Omega_{Ti}-T_{i0}\Omega_{ni}\right]  .
\]

\subsection{The case $\Omega_{ie}=\infty$}

When the frequency $\Omega_{ie}$ is much larger than other frequencies,
$2\Omega_{ie}\gg\partial/\partial t,\Omega_{\chi},\Omega_{Te,i}$, and
$T_{e0}=T_{i0}$, then the dispersion relation becomes the following:%
\begin{equation}
\omega=-\frac{i}{2\gamma}\left(  \Omega_{\chi}+\Omega_{Te}-\Omega_{ne}%
+\Omega_{Ti}-\Omega_{ni}\right)  .
\end{equation}
This is isobaric solution with the electron and ion cooling.

For the different temperatures of electrons and ions, $T_{e0}\neq T_{i0}$, we obtain%

\begin{align}
i\frac{\gamma}{2}\left[  T_{e0}+\left(  4+3\frac{T_{i0}}{T_{e0}}\right)
T_{i0}\right]  \omega &  =\left(  3T_{e0}-T_{i0}\right)  \left(  \Omega_{\chi
}+\Omega_{Te}\right)  -\left[  \frac{5}{2}T_{e0}-3T_{i0}\left(  1+\frac
{T_{i0}}{2T_{e0}}\right)  \right]  \Omega_{Ti}\\
&  -\frac{1}{2}\left(  3T_{i0}+T_{e0}\right)  \left(  \Omega_{ne}+\frac
{T_{i0}}{T_{e0}}\Omega_{ni}\right)  .\nonumber
\end{align}
In the case $T_{e0}\gg T_{i0}$, this equation takes the form,%
\[
\omega=-\frac{i}{\gamma}\left[  6\left(  \Omega_{\chi}+\Omega_{Te}\right)
-\Omega_{ne}-5\Omega_{Ti}\right]  .
\]
In the opposite case, $T_{e0}\ll T_{i0}$, we obtain%
\[
\omega=-\frac{i}{\gamma}\left(  \Omega_{Ti}-\Omega_{ni}\right)  .
\]

\subsection{General case}

In a general case, Equation (48) can be written in the form%
\begin{equation}
g_{0}\omega^{2}+ig_{1}\omega-g_{2}=0,
\end{equation}
where
\begin{align}
g_{0}  &  =\gamma\left(  T_{e0}+T_{i0}\right)  ,\\
g_{1}  &  =\left[  \left(  \gamma T_{i0}+T_{e0}\right)  \left(  \Omega_{\chi
}+\Omega_{Te}\right)  +\left(  \gamma T_{e0}+T_{i0}\right)  \Omega_{Ti}%
-T_{e0}\Omega_{ne}-T_{i0}\Omega_{ni}\right] \nonumber\\
&  +\frac{1}{2}\gamma\left[  T_{e0}+T_{i0}\left(  4+3\frac{T_{i0}}{T_{e0}%
}\right)  \right]  \Omega_{ie},\nonumber\\
g_{2}  &  =T_{e0}\left(  \Omega_{\chi}+\Omega_{Te}-\Omega_{ne}\right)
\Omega_{Ti}+T_{i0}\left(  \Omega_{Ti}-\Omega_{ni}\right)  \left(  \Omega
_{\chi}+\Omega_{Te}\right) \nonumber\\
&  +\left(  3T_{e0}-T_{i0}\right)  \Omega_{ie}\left(  \Omega_{\chi}%
+\Omega_{Te}\right)  -\left[  \frac{5}{2}T_{e0}-3T_{i0}\left(  1+\frac{T_{i0}%
}{2T_{e0}}\right)  \right]  \Omega_{ie}\Omega_{Ti}\nonumber\\
&  -\frac{1}{2}\left(  T_{e0}+3T_{i0}\right)  \Omega_{ie}\left(  \Omega
_{ne}+\Omega_{ni}\frac{T_{i0}}{T_{e0}}\right)  .\nonumber
\end{align}
Equation (52) can be solved numerically for different cooling functions.

\bigskip

\section{DISCUSSION}

From the results obtained above, we can estimate the relative perturbations of
the number density and pressure in the fast sound speed regime. Using
equations (24)-(27) and keeping the main terms, we find expressions for
$\mathbf{\nabla\cdot v}_{j1}$,%
\begin{align}
\frac{\partial}{\partial t}\mathbf{\nabla\cdot v}_{i1} &  \simeq\frac
{\partial}{\partial z}\left(  \alpha_{i}F_{i1z}+\beta_{e}F_{e1z}\right)  ,\\
\frac{\partial}{\partial t}\mathbf{\nabla\cdot v}_{e1} &  \simeq\frac
{\partial}{\partial z}\left(  \alpha_{e}F_{e1z}+\beta_{i}F_{i1z}\right)
.\nonumber
\end{align}
Thus, the number density perturbation $n_{j1}\sim$ $\mathbf{\nabla\cdot
v}_{j1}$ is determined by the longitudinal electric field $E_{z}$. If we use
the condition of quasineutrality, $\mathbf{\nabla\cdot v}_{i1}=\mathbf{\nabla
\cdot v}_{e1}$, and relation $m_{i}\mathbf{F}_{i1}=-m_{e}\mathbf{F}_{e1}$, we
obtain the dispersion relation (42). From equations (21) and (22), taking into
account polarizations (43) and (46), we have%
\begin{equation}
\frac{\partial P_{j1}}{\partial z}\simeq-\frac{\partial F_{j1z}}{\partial t},
\end{equation}
where the value $P_{j1}$ is connected with the pressure perturbation through
the relation%
\begin{equation}
P_{j1}=-\frac{1}{m_{j}n_{j0}}\frac{\partial p_{j1}}{\partial t}.
\end{equation}
From the linear continuity equation and Equations (54)-(56), we find%
\begin{align*}
\frac{\partial^{2}n_{i1}}{\partial t^{2}} &  =-\left(  \frac{\alpha_{i}}%
{m_{i}}-\frac{\beta_{e}}{m_{e}}\right)  \frac{\partial^{2}p_{i1}}{\partial
z^{2}},\\
\frac{\partial^{2}n_{e1}}{\partial t^{2}} &  =-\left(  \frac{\alpha_{e}}%
{m_{e}}-\frac{\beta_{i}}{m_{i}}\right)  \frac{\partial^{2}p_{e1}}{\partial
z^{2}}.
\end{align*}
It is followed from here the relation between $p_{i1}$ and $p_{e1}$,%
\[
\left(  \frac{\alpha_{i}}{m_{i}}-\frac{\beta_{e}}{m_{e}}\right)
p_{i1}=\left(  \frac{\alpha_{e}}{m_{e}}-\frac{\beta_{i}}{m_{i}}\right)
p_{e1}.
\]
The sum of pressures, $p_{e1}+p_{i1}$, is equal to%
\[
p_{e1}+p_{i1}=\delta\left(  \frac{\alpha_{i}}{m_{i}}-\frac{\beta_{e}}{m_{e}%
}\right)  ^{-1}p_{e1}\ll p_{e1}.
\]
Thus, the total pressure almost does not change. An estimation of the value
$\alpha_{i}/m_{i}-\beta_{e}/m_{e}$ gives (see notations (23))%
\[
\left(  \frac{\alpha_{i}}{m_{i}}-\frac{\beta_{e}}{m_{e}}\right)  \sim\frac
{1}{T_{0}}\left(  \frac{\partial}{\partial z}\right)  ^{-2}\frac{\partial^{2}%
}{\partial t^{2}}%
\]
$\left(  T_{e0}\sim T_{i0}\sim T_{0}\right)  $. Thus, we obtain $n_{i1}%
/n_{0}\sim p_{i1}/p_{0}$.

Our dispersion relation (47) does not depend on the wave vector $\mathbf{k}$
(except $\Omega_{\chi}$) and magnetic field $\mathbf{B}_{0}$. This
independence on $\mathbf{k}$ is connected with the limiting case (20). The
absence of the magnetic field is a result of the main contribution to the
dispersion relation for the condensation mode of the longitudinal perturbed
current $j_{1z}$ under the action of the longitudinal electric field $E_{1z}$
(the term $\varepsilon_{zz}E_{1zk}$ in the last equation of the system (40))
which does not contain the magnetic field. Therefore, a wide spectrum of
wavelengths along and across the magnetic field can be generated in the
framework of conditions (18)-(20) and (33). It is worth to note that condition
(20) is a particular case which is considered in this paper. Other conditions
used above are easily satisfied in many astrophysical objects because the ion
cyclotron frequency is much larger than the dynamical and ion-electron
collisional frequencies and the ion Larmor radius is much smaller than the
perturbations wavelengths.

The general form of the dispersion relation (48) including the thermal
exchange between electrons and ions and their different cooling-heating
functions and temperatures permits us to consider various cases which can be
realized in real situations. In particular, Equation (49) is available for a
weak thermal coupling and equal temperatures, while Equations (50) and (51)
are appropriate in the case of strong coupling. The intermediate case is
described by Equation (52), where coefficients (53) contain both different
temperatures and cooling functions. The large thermal conductivity
$\Omega_{\chi}$ stabilizes some perturbations. We note that one particular
Field length (Field 1965) can be obtained from the condition $\Omega_{\chi
}=-\Omega_{Te}+\Omega_{ne}$. However, in the limit $\Omega_{\chi}\gg
\omega,\Omega_{Te,i},\Omega_{ne,i},2\Omega_{ie}$, Equation (48) describes
instability due to the ion cooling (see section 10).

We have shown that unstable perturbations have an electromagnetic nature (see
Equations (43) and (46)). Thus, a consideration of potential perturbations
given in some papers (see section 1) is not adequate.

From a system of equations (40), we see that for condensation mode
$\varepsilon_{zz}=0$ in the fast sound speed regime (20) the longitudinal
electric field $E_{1z}$ plays the crucial role. The transverse wavelengths of
unstable perturbations can be both larger and smaller than the longitudinal ones.

The contribution of collisions between electrons and ions in the momentum
equations depends on the parameter $d$ defined by Equation (38). In both
limits (41) ($d\ll1$) and (44) ($d\gg1$), the dispersion relation has the same
form $\varepsilon_{zz}=0$.

\bigskip

\section{\bigskip ASTROPHYSICAL\ IMPLICATIONS}

The fundamental purpose of this paper is to investigate the thermal
instability taking into account the real multicomponent nature of medium
appropriate, e.g., for galaxy clusters in a straightforward manner, using the
$\mathbf{E}$-approach. Therefore, the different cooling-heating functions for
the electrons and ions, the energy exchange between them, and perturbation of
the energy exchange collision frequency for the non-equilibrium background
state need to be appropriately included. The last effect considered for the
first time for thermal instabilities contribute to the dispersion relation
with the same order of magnitude as other effects and considerably modifies it
(see, e.g., Equations (50) and (51)). We note that the effects mentioned above
are not considered in the astrophysical literature using the MHD-approach. The
key condition for the results obtained is condition (20). It denotes that we
consider sufficienlty short-wavelengths along the magnetic field. In the
simplest case without the energy exchange and ion temperature perturbations,
the growth rate is in the region of the isobaric solution
\[
\omega=-\frac{i}{\gamma}\left(  \Omega_{\chi}+\Omega_{Te}-\Omega_{ne}\right)
\]
(see section 10.1). The same growth rate for the condensation mode, we have,
for example, for the neutral and magnetized media when the perturbation
wavelength tends to infinity (e.g., Field 1965; Heyvaerts 1974; Loewenstein
1990; Balbus 1991). Thus, from Equation (48), we can recover the classical result.

Conditions on the perturbation scale lengths transverse to the magnetic field
are given by inequalities (18), (19), and (33) which can be written in the
form%
\begin{align*}
\frac{k_{\perp}^{2}}{k_{z}^{2}}  & \ll\frac{\omega_{ci}^{2}}{\omega^{2}}%
;\frac{\omega_{ci}^{2}}{\nu_{ie}\omega},\\
k_{\perp}^{2}  & \ll\frac{\omega_{ci}^{2}}{c_{s}^{2}}.
\end{align*}
We see from the first condition that the relation between $k_{\perp}$ and
$k_{z}$ can be arbitrary because $\omega_{ci}^{2}\gg\omega^{2},\nu_{ie}\omega
$, i.e. the cases $k_{\perp}\lesssim k_{z}$ and $k_{\perp}\gg k_{z}$ are
permitted. We emphasize that this conclusion is a consequence of the fast
sound speed limit (20). The case $k_{z}\rightarrow0$ is not suitable for this
limit. The second condition means that the ion sound (or Larmor for
$\Omega_{ie}=0$) radius is much smaller than the transverse wavelength. For
parameters appropriate for galaxy clusters, $B_{0}\sim1$ $\mu$G and $T\sim3$
keV, we have $\omega_{ci}\sim9.6\times10^{-3}$ s$^{-1}$ and $c_{s}\sim\left(
2T/m_{i}\right)  ^{1/2}\sim7.6\times10^{2}$ km s$^{-1}$ (for protons). Thus,
the lower limit to the transverse wavelength $\lambda_{\perp}=2\pi/k_{\perp}$
is equal to $\lambda_{\perp}>5\times10^{5}$ km. We note that since the growth
rates do not depend on $k_{\perp}$ there is no a specific connection between
$k_{\perp}$and $k_{z}$ for the most unstable modes. The upper limit to $k_{z}$
is determind by the thermal conductivity.

The dispersion relation (47) for condensation modes does not contain the
magnetic field because the main dynamics of species in the case under
consideration is along the magnetic field lines and determined by the
longitudinal electric field $E_{z}$. This field is not captured by the ideal
magnetic induction equation. We have obtained the dispersion relation which
cannot followed from the MHD consideration. At the same time, the simplest
condensation mode growth rate described by Equation (47) coincides with the
MHD result (see above this section).

We further shortly outline some important points of our investigation for
possible observations. We have found the growth rates which contain the
separate contribution of the cooling functions of electrons and ions in a
clear form. It is obvious that both components (in multicomponent media, the
dust grains and neutrals can also be present) can result in the thermal
instability in the similar manner. This fact considerably extends
possibilities for a medium to become unstable. In this connection, it is
important to know the functional dependence of the cooling functions on the
temperature and density for each species. The range of the scale lengths of
unstable perturbations can enlarge due to contribution of other species except
electrons to the instability. For example, the short-wavelength perturbations
which must be stable because of the large electron thermal conduction can be
unstable due to contribution of ions to cooling of medium (see section 10). In
the fast sound regime when the growth rates do not depend on the magnetic
field and wavelengths of perturbations, both filaments and clouds and pancakes
can be observed. It is important to have in mind that in this regime the
electric field and electric currents of species along the magnetic field
lines\textbf{\ }have the crucial role. This result could be used for the
diagnostics of the magnetic field direction in unstable domains.

\bigskip

\section{\bigskip CONCLUSION}

We have treated the thermal instability in the electron-ion magnetized plasma
which is relevant to galaxy clusters, solar corona, and other two-component
plasma objects. The multicomponent plasma approach have been applied to derive
the dispersion relation for the condensation modes in the case in which the
dynamical frequency is much slower than the sound frequency. Our dispersion
relation takes into account the electron and ion cooling-heating functions,
collisions in the momentum equations, energy exchange in the thermal
equations, different background temperatures of the electrons and ions, and
perturbation of the energy exchange frequency due to density and temperature
perturbations. Different limiting cases of the dispersion relation have been
considered and simple expressions for the growth rates have been obtained. We
have shown that perturbations have an electromagnetic nature. The important
role of the electric field perturbation along the background magnetic field
has been demonstrated. We have found that at conditions under consideration,
the condensation must occur along the magnetic field lines while the
transverse scale sizes can be both larger and smaller than the longitudinal
ones. General expressions for the dynamical variables obtained in this paper
can be applied for a wide range of astrophysical and laboratory plasmas also
containing the neutrals and dust grains. The results obtained can be useful
for interpretation of observations of the dense cold regions in astrophysical
objects such as IGM, solar corona, and so on.

\bigskip

I would like gratefully to acknowledge Dr. Mohsen Shadmehri for valuable
discussions and suggestions and also the anonymous referee, whose useful comments helped in improving the manuscript.

\bigskip


\begin{appendix}
\section{APPENDIX}
\subsection{Perturbed velocities of species}
In the linear approximation, Equation (1) for the perturbed velocity
$\mathbf{v}_{j1}$ takes the form%
\begin{equation}
\frac{\partial\mathbf{v}_{j1}}{\partial t}=-\frac{\mathbf{\nabla}p_{j1}}%
{m_{j}n_{j0}}+\mathbf{F}_{j1}\mathbf{+}\frac{q_{j}}{m_{j}c}\mathbf{v}%
_{j1}\times\mathbf{B}_{0},\\ 
\end{equation}
where $p_{j1}=n_{j0}T_{j1}+n_{j1}T_{j0}$. From this equation, we can find
solutions for the components of $\mathbf{v}_{j1}$. For simplicity, we assume
that $\partial/\partial x=0$ because a system is symmetric in the transverse
direction relative to the $z$-axis. Then, the $x$-component of Equation (A1)
gives%
\begin{equation}
\frac{\partial v_{j1x}}{\partial t}=F_{j1x}\mathbf{+}\omega_{cj}%
v_{j1y},\\ 
\end{equation}
where $\omega_{cj}=q_{j}B_{0}/m_{j}c$ is the cyclotron frequency of species
$j$. Differentiating Equation (A1) over $t$ and using Equation (2) in the
linear approximation and Equations (15), (16), and (A2), we obtain for the $y
$-component of Equation (A1)%
\begin{equation}
\left(  \frac{\partial^{2}}{\partial t^{2}}+\omega_{cj}^{2}\right)
v_{j1y}=\frac{\partial P_{j1}}{\partial y}+Q_{j1y},\\ 
\end{equation}
where%
\begin{align}
P_{e1}  & =-\frac{G_{2}}{Dm_{e}}\frac{\partial}{\partial t}\mathbf{\nabla
}\cdot\mathbf{v}_{i1}+\left(  \frac{T_{e0}}{m_{e}}-\frac{G_{1}}{Dm_{e}}%
\frac{\partial}{\partial t}\right)  \ \mathbf{\nabla}\cdot\mathbf{v}%
_{e1},\\ 
P_{i1}  & =-\frac{G_{3}}{Dm_{i}}\frac{\partial}{\partial t}\mathbf{\nabla
}\cdot\mathbf{v}_{e1}+\left(  \frac{T_{i0}}{m_{i}}-\frac{G_{4}}{Dm_{i}}%
\frac{\partial}{\partial t}\right)  \mathbf{\nabla}\cdot\mathbf{v}%
_{i1}.\nonumber
\end{align}
The value $P_{j1}$ is connected with the pressure perturbation (see Equation
(A1)). Using Equations (A2) and (A3), we find
\begin{equation}
\frac{\partial}{\omega_{cj}\partial t}\left[  \left(  \frac{\partial^{2}%
}{\partial t^{2}}+\omega_{cj}^{2}\right)  v_{j1x}-Q_{j1x}\right]
\mathbf{=}\frac{\partial P_{j1}}{\partial y}\\ 
\end{equation}
In Equations (A3) and (A5), notations%
\begin{align}
Q_{j1y}  & =-\omega_{cj}F_{j1x}+\frac{\partial F_{j1y}}{\partial t},\\ 
Q_{j1x}  & =\omega_{cj}F_{j1y}\mathbf{+}\frac{\partial F_{j1x}}{\partial
t}\nonumber
\end{align}
are introduced. We see from these equations that the thermal pressure effect
on the velocity $v_{i1x}$ is much larger than that on $v_{i1y}$ when
$\partial/\partial t\ll\omega_{ci}$. The $z$-component of Equation (A1) can be
written in the form%
\begin{equation}
\frac{\partial^{2}v_{j1z}}{\partial t^{2}}=\frac{\partial P_{j1}}{\partial
z}+\frac{\partial F_{j1z}}{\partial t}.\\ 
\end{equation}
\subsection{Calculation of $\mathbf{\nabla\cdot v}_{j1}$ and $P_{j1} $}
We have%
\begin{equation}
\mathbf{\nabla}\cdot\mathbf{v}_{j1}=\frac{\partial v_{j1y}}{\partial y}%
+\frac{\partial v_{j1z}}{\partial z}.\\ 
\end{equation}
Using Equations (A3), (A4), (A7) and (A8), we obtain%
\begin{align}
L_{1e}\ \mathbf{\nabla}\cdot\mathbf{v}_{e1}+L_{2e}\mathbf{\nabla}%
\cdot\mathbf{v}_{i1}  & =H_{e1},\\ 
L_{1i}\mathbf{\nabla}\cdot\mathbf{v}_{i1}+L_{2i}\mathbf{\nabla}\cdot
\mathbf{v}_{e1}  & =H_{i1}.\nonumber
\end{align}
Here,%
\begin{equation}
H_{j1}=\frac{\partial^{3}Q_{j1y}}{\partial y\partial t^{2}}+\left(
\frac{\partial^{2}}{\partial t^{2}}+\omega_{cj}^{2}\right)  \frac{\partial
^{2}F_{j1z}}{\partial z\partial t}\\ 
\end{equation}
and operators $L_{1j}$ and $L_{2j}$ are the following:%
\begin{align}
L_{1e}  & =\left(  \frac{\partial^{2}}{\partial t^{2}}+\omega_{ce}^{2}\right)
\frac{\partial^{2}}{\partial t^{2}}-L_{3e}\left(  \frac{T_{e0}}{m_{e}}%
-\frac{G_{1}}{Dm_{e}}\frac{\partial}{\partial t}\right)  ,\\ 
L_{1i}  & =\left(  \frac{\partial^{2}}{\partial t^{2}}+\omega_{ci}^{2}\right)
\frac{\partial^{2}}{\partial t^{2}}-L_{3i}\left(  \frac{T_{i0}}{m_{i}}%
-\frac{G_{4}}{Dm_{i}}\frac{\partial}{\partial t}\right)  ,\nonumber\\
L_{2e}  & =L_{3e}\frac{G_{2}}{Dm_{e}}\frac{\partial}{\partial t},L_{2i}%
=L_{3i}\frac{G_{3}}{Dm_{i}}\frac{\partial}{\partial t},\nonumber\\
L_{3j}  & =\frac{\partial^{4}}{\partial y^{2}\partial t^{2}}\ +\left(
\frac{\partial^{2}}{\partial t^{2}}+\omega_{cj}^{2}\right)  \frac{\partial
^{2}}{\partial z^{2}}.\nonumber
\end{align}
From a system of equations (A9), we find
\begin{align}
L\mathbf{\nabla}\cdot\mathbf{v}_{e1}  & =-L_{2e}H_{i1}+L_{1i}H_{e1}%
,\\ 
L\mathbf{\nabla}\cdot\mathbf{v}_{i1}  & =-L_{2i}H_{e1}+L_{1e}\ H_{i1}%
,\nonumber
\end{align}
where
\begin{equation}
L=L_{1e}L_{1i}\ -L_{2e}L_{2i}.\\ 
\end{equation}
The values $P_{i1}$ and $P_{e1}$ can be found, substituting solutions (A12)
into expressions (A4),%
\begin{align}
LP_{i1}  & =\left[  \frac{G_{3}}{Dm_{i}}\frac{\partial}{\partial t}%
L_{2e}+\left(  \frac{T_{i0}}{m_{i}}-\frac{G_{4}}{Dm_{i}}\frac{\partial
}{\partial t}\right)  L_{1e}\right]  \ H_{i1}\\ 
& -\left[  \frac{G_{3}}{Dm_{i}}\frac{\partial}{\partial t}L_{1i}+\left(
\frac{T_{i0}}{m_{i}}-\frac{G_{4}}{Dm_{i}}\frac{\partial}{\partial t}\right)
L_{2i}\right]  H_{e1},\nonumber
\end{align}%
\begin{align}
LP_{e1}  & =\left[  \frac{G_{2}}{Dm_{e}}\frac{\partial}{\partial t}%
L_{2i}+\left(  \frac{T_{e0}}{m_{e}}-\frac{G_{1}}{Dm_{e}}\frac{\partial
}{\partial t}\right)  L_{1i}\right]  H_{e1}\\ 
& -\left[  \frac{G_{2}}{Dm_{e}}\frac{\partial}{\partial t}L_{1e}\ +\left(
\frac{T_{e0}}{m_{e}}-\frac{G_{1}}{Dm_{e}}\frac{\partial}{\partial t}\right)
L_{2e}\right]  H_{i1}.\nonumber
\end{align}
\subsection{Expressions for $D$ and $G_{1,2,3,4}$}
We now give expressions for values given by a system (17):
\begin{equation}
D=\left(  \frac{\partial}{\partial t}+\Omega_{\chi}+\Omega_{Te}\right)
\left(  \frac{\partial}{\partial t}+\Omega_{Ti}+\Omega_{ie}\right)
\frac{\partial^{2}}{\partial t^{2}}+\left(  \Omega_{ei}+\Omega_{Tei}\right)
\left(  \frac{\partial}{\partial t}+\Omega_{Ti}\right)  \frac{\partial^{2}%
}{\partial t^{2}},\\ 
\end{equation}%
\begin{equation}
G_{1}=T_{e0}\left[  \Omega_{ne}-\left(  \gamma-1\right)  \frac{\partial
}{\partial t}\right]  \left(  \frac{\partial}{\partial t}+\Omega_{Ti}%
+\Omega_{ie}\right)  \frac{\partial}{\partial t}+\Omega_{ei}\left(
T_{e0}-T_{i0}\right)  \left(  \frac{\partial}{\partial t}+\Omega_{Ti}\right)
\frac{\partial}{\partial t},\\ 
\end{equation}%
\begin{equation}
G_{2}=\Omega_{ei}T_{i0}\left[  \Omega_{ni}-\left(  \gamma-1\right)
\frac{\partial}{\partial t}\right]  \frac{\partial}{\partial t}+\Omega
_{ei}\left(  T_{e0}-T_{i0}\right)  \left(  \frac{\partial}{\partial t}%
+\Omega_{Ti}\right)  \frac{\partial}{\partial t},\\ 
\end{equation}%
\begin{equation}
G_{3}=\left(  \Omega_{Tie}+\Omega_{ie}\right)  T_{e0}\left[  \Omega
_{ne}-\left(  \gamma-1\right)  \frac{\partial}{\partial t}\right]
\frac{\partial}{\partial t}-\Omega_{ie}\left(  T_{e0}-T_{i0}\right)  \left(
\frac{\partial}{\partial t}+\Omega_{\chi}+\Omega_{Te}\right)  \frac{\partial
}{\partial t},\\ 
\end{equation}%
\begin{align}
G_{4}  & =T_{i0}\left(  \frac{\partial}{\partial t}+\Omega_{\chi}+\Omega
_{Te}+\Omega_{Tei}+\Omega_{ei}\right)  \left[  \Omega_{ni}-\left(
\gamma-1\right)  \frac{\partial}{\partial t}\right]  \frac{\partial}{\partial
t}\\ 
& -\Omega_{ie}\left(  T_{e0}-T_{i0}\right)  \left(  \frac{\partial}{\partial
t}+\Omega_{\chi}+\Omega_{Te}\right)  \frac{\partial}{\partial t}.\nonumber
\end{align}
\subsection{Simplification of Equations (A14) and (A15)}
We shall further calculate coefficients by $H_{j1}$ in Equations (A14) and
(A15). Using expressions (A11), we find
\begin{equation}
\frac{G_{3}}{Dm_{i}}\frac{\partial}{\partial t}L_{1i}+\left(  \frac{T_{i0}%
}{m_{i}}-\frac{G_{4}}{Dm_{i}}\frac{\partial}{\partial t}\right)  L_{2i}%
=\frac{G_{3}}{Dm_{i}}\left(  \frac{\partial^{2}}{\partial t^{2}}+\omega
_{ci}^{2}\right)  \frac{\partial^{3}}{\partial t^{3}}\\ 
\end{equation}
and%
\begin{equation}
\frac{G_{3}}{Dm_{i}}\frac{\partial}{\partial t}L_{2e}+\left(  \frac{T_{i0}%
}{m_{i}}-\frac{G_{4}}{Dm_{i}}\frac{\partial}{\partial t}\right)  L_{1e}%
=\frac{1}{D}\left(  D\frac{T_{i0}}{m_{i}}-\frac{G_{4}}{m_{i}}\frac{\partial
}{\partial t}\right)  \left(  \frac{\partial^{2}}{\partial t^{2}}+\omega
_{ce}^{2}\right)  \frac{\partial^{2}}{\partial t^{2}}+\frac{1}{Dm_{e}m_{i}%
}L_{3e}K.\\ 
\end{equation}
In Equation (A22), we have introduced notation%
\begin{equation}
K=\frac{1}{D}\left(  G_{2}G_{3}-G_{1}G_{4}\right)  \frac{\partial^{2}%
}{\partial t^{2}}+\left(  T_{e0}G_{4}+T_{i0}G_{1}\right)  \frac{\partial
}{\partial t}-DT_{e0}T_{i0}.\\ 
\end{equation}
Calculations show that the value $\left(  G_{2}G_{3}-G_{1}G_{4}\right)  $ has
a simple form, i.e.%
\begin{align}
\frac{1}{D}\left(  G_{2}G_{3}-G_{1}G_{4}\right)   & =\Omega_{ie}\left(
T_{e0}-T_{i0}\right)  T_{e0}\left[  \Omega_{ne}-\left(  \gamma-1\right)
\frac{\partial}{\partial t}\right]  +\Omega_{ei}\left(  T_{i0}-T_{e0}\right)
T_{i0}\left[  \Omega_{ni}-\left(  \gamma-1\right)  \frac{\partial}{\partial
t}\right] \\ 
& -T_{e0}T_{i0}\left[  \Omega_{ne}-\left(  \gamma-1\right)  \frac{\partial
}{\partial t}\right]  \left[  \Omega_{ni}-\left(  \gamma-1\right)
\frac{\partial}{\partial t}\right]  .\nonumber
\end{align}
Using expressions (A16), (A17), (A20), and (A24), we obtain for the operator
$K$ (A23) the simple form,
\begin{equation}
K=-\Omega_{ie}T_{e0}^{2}W_{e}\frac{\partial^{2}}{\partial t^{2}}-\left(
\Omega_{ei}T_{i0}+\Omega_{Tei}T_{e0}\right)  T_{i0}W_{i}\frac{\partial^{2}%
}{\partial t^{2}}-T_{e0}T_{i0}W_{e}W_{i}\frac{\partial^{2}}{\partial t^{2}%
},\\ 
\end{equation}
where notations
\begin{align}
W_{e}  & =\gamma\frac{\partial}{\partial t}+\Omega_{\chi}+\Omega_{Te}%
-\Omega_{ne},\\ 
W_{i}  & =\gamma\frac{\partial}{\partial t}+\Omega_{Ti}-\Omega_{ni}\nonumber
\end{align}
are introduced. Using Equations (A21) and (A22), Equation (A14) for $P_{i1}$
takes the form%
\begin{align}
DLP_{i1}  & =\left[  \left(  D\frac{T_{i0}}{m_{i}}-\frac{G_{4}}{m_{i}}%
\frac{\partial}{\partial t}\right)  \left(  \frac{\partial^{2}}{\partial
t^{2}}+\omega_{ce}^{2}\right)  \frac{\partial^{2}}{\partial t^{2}}+\frac
{1}{m_{e}m_{i}}L_{3e}K\right]  \ H_{i1}\\ 
& -\frac{G_{3}}{m_{i}}\left(  \frac{\partial^{2}}{\partial t^{2}}+\omega
_{ci}^{2}\right)  \frac{\partial^{3}}{\partial t^{3}}H_{e1}.\nonumber
\end{align}
Analogous consideration of Equation (A15) leads to the following equation for
$P_{e1}$:
\begin{align}
DLP_{e1}  & =\left[  \left(  D\frac{T_{e0}}{m_{e}}-\frac{G_{1}}{m_{e}}%
\frac{\partial}{\partial t}\right)  \left(  \frac{\partial^{2}}{\partial
t^{2}}+\omega_{ci}^{2}\right)  \frac{\partial^{2}}{\partial t^{2}}+\frac
{1}{m_{i}m_{e}}L_{3i}K\right]  H_{e1}\\ 
& -\frac{G_{2}}{m_{e}}\left(  \frac{\partial^{2}}{\partial t^{2}}+\omega
_{ce}^{2}\right)  \frac{\partial^{3}}{\partial t^{3}}H_{i1}.\nonumber
\end{align}
Operators
\begin{align*}
& D\frac{T_{e0}}{m_{e}}-\frac{G_{1}}{m_{e}}\frac{\partial}{\partial t},\\
& D\frac{T_{i0}}{m_{i}}-\frac{G_{4}}{m_{i}}\frac{\partial}{\partial t}%
\end{align*}
can be found by using Equations (A16), (A17), (A20), and (A26)%
\begin{align}
D\frac{T_{e0}}{m_{e}}-\frac{G_{1}}{m_{e}}\frac{\partial}{\partial t}  &
=\frac{T_{e0}}{m_{e}}W_{e}\left(  \frac{\partial}{\partial t}+\Omega
_{Ti}+\Omega_{ie}\right)  \frac{\partial^{2}}{\partial t^{2}}\\ 
& +\frac{1}{m_{e}}\left(  T_{e0}\Omega_{Tei}+\Omega_{ei}T_{i0}\right)  \left(
\frac{\partial}{\partial t}+\Omega_{Ti}\right)  \frac{\partial^{2}}{\partial
t^{2}},\nonumber\\
D\frac{T_{i0}}{m_{i}}-\frac{G_{4}}{m_{i}}\frac{\partial}{\partial t}  &
=\frac{T_{i0}}{m_{i}}W_{i}\left(  \frac{\partial}{\partial t}+\Omega_{\chi
}+\Omega_{Te}+\Omega_{ei}+\Omega_{Tei}\right)  \frac{\partial^{2}}{\partial
t^{2}}\nonumber\\
& +\frac{T_{e0}}{m_{i}}\Omega_{ie}\left(  \frac{\partial}{\partial t}%
+\Omega_{\chi}+\Omega_{Te}\right)  \frac{\partial^{2}}{\partial t^{2}%
}.\nonumber
\end{align}
\subsection{Operator $L$ in a general form}
Using expressions (A11), we find from Equation (A13)%
\begin{equation}
L=M-N-\frac{1}{m_{e}m_{i}D}L_{3e}L_{3i}K,\\ 
\end{equation}
where
\begin{align}
M  & =\left(  \frac{\partial^{2}}{\partial t^{2}}+\omega_{ce}^{2}\right)
\left(  \frac{\partial^{2}}{\partial t^{2}}+\omega_{ci}^{2}\right)
\frac{\partial^{4}}{\partial t^{4}},\\ 
N  & =\left(  \frac{\partial^{2}}{\partial t^{2}}+\omega_{ci}^{2}\right)
\frac{\partial^{2}}{\partial t^{2}}L_{3e}\left(  \frac{T_{e0}}{m_{e}}%
-\frac{G_{1}}{Dm_{e}}\frac{\partial}{\partial t}\right)  +\left(
\frac{\partial^{2}}{\partial t^{2}}+\omega_{ce}^{2}\right)  \frac{\partial
^{2}}{\partial t^{2}}L_{3i}\left(  \frac{T_{i0}}{m_{i}}-\frac{G_{4}}{Dm_{i}%
}\frac{\partial}{\partial t}\right)  .\nonumber
\end{align}
\end{appendix}

\bigskip

\section{\bigskip REFERENCES}

Audit, E., \& Hennebelle, P. 2005, A\&A, 433,1

Balbus, S. A. 1986, ApJ, 303, L79

Balbus, S. A. 1991, ApJ, 372, 25

Balbus, S. A., \& Soker, N. 1989, ApJ, 341, 611

Begelman, M. C., \& McKee, C. F. 1990, ApJ, 358, 375

Begelman, M. C., \& Zweibel, E. G. 1994, ApJ, 431, 689

Birk, G. T. 2000, Phys. Plasmas, 7, 3811

Birk, G. T., \& Wiechen, H. 2001, Phys. Plasmas, 8, 5057

Bogdanovi\'{c}, T., Reynolds, C. S., Balbus, S. A., \& Parrish, I. J. 2009,
ApJ, 704, 211

Bora, M. P., \& Taiwar, S. P. 1993, Phys. Fluids B, 5, 950

Braginskii, S. I. 1965, Rev. Plasma Phys., 1, 205

Burkert, A., \& Lin, D. N. C. 2000, ApJ, 537, 270

Cox, D. P. 2005, ARA\&A, 43, 337

Elmegreen, B. G., \& Scalo, J. 2004, ARA\&A, 42, 211

Ettori, S. \& Fabian, A. C. 1998, MNRAS, 293, L33

Fox, D. C., \& Loeb, A. 1997, ApJ, 491, 459

Fukue, T., \& Kamaya, H. 2007, ApJ, 669, 363

Field, G.B. 1965, ApJ, 142, 531

Hennebelle, P., \& Audit, E. 2007, A\&A, 465, 431

Hennebelle, P., \& P\'{e}rault, M. 1999, A\&A, 351, 309

Hennebelle, P., \& P\'{e}rault, M. 2000, A\&A, 359, 1124

Heyvaerts, J. 1974, A\&A, 37, 65

Heiles, C., \& Crutcher, R. 2005, in Cosmic Magnetic Fields, ed. R.Wielebinski

\& R. Beck (Lecture Notes in Physics) (Berlin: Springer)

Ib\'{a}\~{n}ez, M. H., \& Shchekinov, Yu. A. 2002, Phys. Plasmas, 9, 3259

Inoue, T., \& Inutsuka, S. 2008, ApJ, 687, 303

Gomez-Pelaez, A. J. \& Moreno-Insertis, F. 2002, ApJ, 569, 766

Karpen, J. T., Picone, J. M., \& Dahlburg, R. B. 1988, ApJ, 324, 590

Karpen, J. T., Antiochos, S. K., Picone, J. M., \& Dahlburg, R. B. 1989, ApJ,
338, 493

Kopp, A., \& Shchekinov, Yu. A. 2007, Phys. Plasmas, 14, 073701

Kopp, A., Schr\"{o}er, A., Birk, G.T., \& Shukla, P. K. 1997, Phys. Plasmas,
4, 4414

Koyama, H., \& Inutsuka, S. 2000, ApJ, 532, 980

Koyama, H., \& Inutsuka, S. 2002, ApJ, 564, L97

Kritsuk, A. G., \& Norman, M. L. 2002, ApJ, 569, L127

Loewenstein, M. 1990, ApJ, 349, 471

Markevitch M., Mushotzky R., Inoue H., Yamashita K., Furuzawa A.,

Tawara Y., 1996, ApJ, 456, 437

Mason, S. F., \& Bessey, R.J. 1983, Solar Phys., 83, 121

Mathews, W., \& Bregman, J. 1978, ApJ, 224, 308

Meerson, B. 1996, Rev. Mod. Phys., 68, 215

Nakagawa, Y. 1970, Sol. Phys., 12, 419

Nekrasov, A. K. 2009 a, ApJ, 695, 46

Nekrasov, A. K. 2009 b, ApJ, 704, 80

Nekrasov, A. K. 2009 c, MNRAS, 400, 1574

Nekrasov, A. K., \& Shadmehri, M. 2010, ApJ, 724,\textbf{\ }1165

Nekrasov, A. K., \& Shadmehri, M. 2011, Astrophys. Space Sci. (in press)

Nipoti, C. 2010, MNRAS, 406, 247

Pandey, B. P., \& Krishan, V. 2001, IEEE Trans. Plasma Sci., 29, 307

Pandey, B. P., Vranje\v{s}, J., \& Parshi, S. 2003, Pramana, 60, 491

Parker, E. N. 1953, ApJ, 117, 431

Parrish, I. J., Quataert, E., \& Sharma, P. 2009, ApJ., 703, 96

S\'{a}nchez-Salcedo, F. J., V\'{a}zquez-Semadeni, E., \& Gazol, A. 2002, ApJ,
577, 768

Shadmehri, M., Nejad-Asghar, M., \& Khesali, A. 2010, Ap\&SS, 326, 83

Sharma, P., Parrish, I. J., \& Quataert, E. 2010, ApJ, 720, 652

Shukla, P. K., \& Sandberg, I. 2003, Phys. Rev. E, 67, 036401

Stiele, H., Lesch, H., \& Heitsch, F. 2006, MNRAS, 372, 862

Takizawa, M. 1998, ApJ, 509, 579

Tozzi, P., \& Norman, C. 2001, ApJ, 546, 63

Trevisan, M. C., \& Ib\'{a}\~{n}ez, M. H. 2000, Phys. Plasmas, 7, 897

Yatou, H., \& Toh, S. 2009, Phys. Rev. E, 036314

V\'{a}zquez-Semadeni, E., Gazol, A., S%
\'{}%
anchez-Salcedo, F. J., \& Passot, T. 2003, Lecture Notes in Physics 614 ed. T.
Passot \& E. Falgarone, 213

V\'{a}zquez-Semadeni, E., Ryu, D., Passot, T., Gonz\'{a}lez, R. F., \& Gazol,
A. 2006, ApJ, 643, 245

\end{document}